# Power Cyber-Physical System Risk Area Prediction Using Dependent Markov Chain and Improved Grey Wolf Optimization

ZHAOYANG QU[1,2], QIANHUI XIE[1], YUQING LIU[1,3], YANG LI[4], LEI WANG[1], PENGCHENG XU[5], YUGUANG ZHOU[5], JIAN SUN[5], KAI XUE[5], AND MINGSHI CUI[6]

[1]College of Information Engineering, Northeast Electric Power University, Jilin 132012, China
[2]Jilin Engineering Technology Tesearch Center of Intelligent Electric Power Big Data Processing, Jilin 132012, China
[3]Department of Electrical and Electronic Engineering, University of Bath, Bath, UK
[4]School of Electrical Engineering, Northeast Electric Power University, Jilin 132012, China
[5] State Grid Jilin Electric Power Supply Company, Changchun 13000, China
[6]State Grid Neimenggu Eastern Electric Power Supply Company, Huhehaote 010000, China

Corresponding author: Qianhui Xie (e-mail: xieqianhui@ yeah.net).

ORCID：Qianhui Xie, https://orcid.org/0000-0002-6410-2941

This work was supported in part by the Key Projects of the National Natural Science Foundation of China under Grant 51437003,and in part by the Jilin Science and Technology Development Plan Project of China under Grant 20160623004T and Grant 20180201092GX.

**ABSTRACT** Existing power cyber-physical system (CPS) risk prediction results are inaccurate as they fail to reflect the actual physical characteristics of the components and the specific operational status. A new method based on dependent Markov chain for power CPS risk area prediction is proposed in this paper. The load and constraints of the non-uniform power CPS coupling network are first characterized, and can be utilized as a node state judgment standard. Considering the component node isomerism and interdependence between the coupled networks, a power CPS risk regional prediction model based on dependent Markov chain is then constructed. A cross-adaptive gray wolf optimization algorithm improved by adaptive position adjustment strategy and cross-optimal solution strategy is subsequently developed to optimize the prediction model. Simulation results using the IEEE 39-BA 110 test system verify the effectiveness and superiority of the proposed method.

## I. INTRODUCTION
### A. BACKGROUND

The development of an increasing number of varied energy sources requires the traditional power system to connect with a variety of cyber systems such as computing equipment, communication devices, and sensors. As a result, the cyber-physical power system has become increasingly complex and multi-dimensional [1]. This tight interconnection not only provides support for data analysis and intelligent control decisions by power systems, but also means the grid is more dependent on the cyber network. The security of cyber systems therefore significantly affects the stable operation of the power system [2]. Attacks on cyber networks have caused large-scale power outages in numerous incidents at home and abroad, such as the cases detailed in [3]-[5]. Therefore, it is of great importance to reduce the system risk and improve the stability of the cyber-physical system (CPS) by simulating the risk propagation process and predicting the possible affected risk area promptly and accurately after the cyber network is attacked [6].

At present, two aspects are the main focus of research into the risk propagation process and prediction method of the cyber physical system. Firstly, graph theory and percolation theory are utilized to analyze the risk propagation and development process from the perspective of physical characteristics such as network topology. Reference [7] analyzed the risk propagation process in power systems of small-world effects. It was found that the risk was concentrated in the high clustering area, and a strategy was developed to mitigate the damage of network faults. The risk propagation process was explored in [8] by adjusting network topologies such as node degree and degree distribution of the network. The authors in [9] used the method of cellular automata to compare the power information physical system to physical cells and information cells, and established the security based on cellular automata through the relationship between the normal and fault states of the two cells. The system risk propagation process was simulated from the perspective of characteristics of load flow operation in the system in [10]-[13]. The real-time operating state of the system was then analyzed based on the node load optimization reconfiguration process and the impact of the cyber routing strategy, power network load flow calculation, and coupled network on risk diffusion and propagation was explored. A risk prediction method was then proposed based on complex networks. The authors considered the functions undertaken by the power information system, using the monitoring and control failure model to evaluate the reliability of the power information physical system in [14]-[15].

### B. RESEARCH MOTIVATION
Although a large body of research has been conducted on power CPS risk prediction, some limitations remain and are detailed in this subsection. A lot of specific operation processes of the power cyber-physical system are simplified in research of the propagation process based on the network topology structure, which ignores the coupling relationship and interaction influence between the power network and the cyber state. The propagation trend of power CPS risk cannot be accurately simulated by simply coupling the cyber node and the power node in a one-to-one manner. Most existing research based on system load flow operation focuses on the overall grasp of load flow characteristics, often ignoring the specific physical properties of different component nodes and their real-time status information due to the excessively large and complex structural characteristics of power cyber-physical systems. As a result, the prediction results are inaccurate.

### C. RESEARCH CONTRIBUTIONS
The main contributions of this paper are outlined as follows.
- A risk area prediction model for power CPS is formulated by considering the effects of load reconfiguration between nodes in the same network and the interdependence between the coupled networks. This model is more accurately reflects the actual system risk propagation process.
- A cross-adaptive grey wolf optimization based on crossover strategy is proposed. Adaptive degree position adjustment strategy and cross optimal solution strategy are then introduced to enhance the efficiency and accuracy of the grey wolf optimization result.

### D. ORGANIZATION
The remainder of this paper is organized as follows. The load and constraints of the non-uniform power CPS coupling network are formally characterized in Section II. In Section III, the risk area prediction model for power CPS is constructed, which can distinguish different node properties and capture the interdependence between the coupled networks. The proposed cross-adaptive gray wolf optimization algorithm (CAGWO) is presented in Section IV, and the simulation experiment is explained and results are outlined and discussed in Section V. Finally, conclusions are drawn in Section VI.

## II. CHARACTERIZATION OF NETWORK LOAD AND CONSTRAINTS OF NON-UNIFORM POWER CPS

As each node in the real power cyber-physical system carries loads with actual physical meaning, the cyber node is responsible for transmitting the information flow and the physical network node is responsible for transmitting the current. The loads carried by these nodes are related to its elements such as location and material properties, and it is difficult to determine the actual node load for research and simulation. Therefore, in the research of this paper, the load model calculated by combining the network topology and the operating characteristics is employed to replace the real load. Additionally, the improved ball warehouse model is used to non-uniformly distribute the coupling part in order to achieve formal characterization of the physical fusion of power information.

### A. CYBER FLOW LOAD AND CONSTRAINT CHARACTERIZATION

A single cyber is represented by $G_c=(V_c,E_c)$, where $V_c$ represents a set of nodes and $E_c$ represents a set of edges. The total load of the cyber per unit time can be described as:

$$T_n = \sum_{j \in V_{p-normal}} \alpha \cdot l_{nj}^{\delta} \qquad (1)$$

Among them, it is believed that the faulty physical network node is not monitored and controlled by the cyber node, so $V_{p-normal}$ is the normal working set in the physical network, $l_{nj}$ is the node degree of the physical network node $j$ in time step $n$, and $\alpha$ and $\delta$ are random variables used to control the load of the cyber to the normal working node in the physical network. The cyber flow load only includes the structural load of the cyber, while specific service classifications are ignored.

The load $LC_{nj}$ of the cyber node $j$ in time step $n$ is:

$$L_{nj}^{c} = \frac{T_n \cdot d_{nj}^{\theta}}{\sum_{j \in V_{c-n0\ nomal}} d_{nj}^{\theta}} \qquad (2)$$

where $V_{p-normal}$ is the normal working node in the cyber, $d_{ni}$ is the node degree of the cyber node $j$ in the time step $n$, and $\theta$ is the influence factor of the cyber node bearing the load. When $\theta>0$, the bigger the node degree, the greater the load it bears.

At the same time, because the load that the cyber node can bear is limited, once a cyber node is attacked, a large number of data streams will flow into the unified node, causing the buffer of the node device to overflow and packet loss. At this time, the cyber node cannot transmit the cyber stream normally. The constraint of the cyber node is:

$$L_{nj}^{C} \leq (1+\rho_c) \cdot L_{0j}^{C} \quad (3)$$

where $\rho_c$ is the tolerance coefficient of the cyber node and $LC_{0j}$ is the load connected to the node $j$ before system failure.

### B. PHYSICAL NETWORK LOAD AND CONSTRAINT CHARACTERIZATION

A single power network is represented by $G_c=(V_c,E_c)$, where $V_p$ represents a set of nodes and $E_p$ represents a set of edges. There are three types of nodes in a physical network: a power generation node that produces electrical energy, a load node that consumes electrical energy, and a substation node that neither produces nor consumes electrical energy. As the power consumption of the cyber node is much smaller than the total power consumption of the entire network, the impact of the cyber on the power network load can be ignored.

The load $LP_{nh}$ of the physical network node $h$ in time step $n$ is:

$$L_{nh}^{p} = \beta \cdot l_{0h}^{\mu} \quad (4)$$

where $\beta$ and $\mu$ are random variables and $l_{0h}$ is the node degree of the physical network node $h$ at the initial moment. Similarly, as a physical network node transmitting electrical energy, once a fault occurs, the loads carried by it are redistributed and transferred to its neighbor nodes. During this period, one or more nodes may bear more load than their capacity and thus cannot transmit power normally. The constraint of physical network nodes is:

$$L_{nh}^{p} \leq (1+\rho_p) \cdot \sqrt{L_{0h}^{p} \cdot P_{nh}} \quad (5)$$

### C. ESTABLISHMENT CYBER-PHYSICAL COUPLING OF NON-UNIFORM INTERDEPENDENCE

The nodes in the cyber network are divided into monitoring function nodes and control function nodes according to the "k-n" model [16]. The control node performs calculation and generates operation of the cyber system directly by monitoring the cyber system uploaded by the node to the physical network, and each node in the physical network is governed by the control node. At the same time, all nodes in the cyber network can also obtain power from the physical network. The coupling correspondence is shown in Fig. 1.

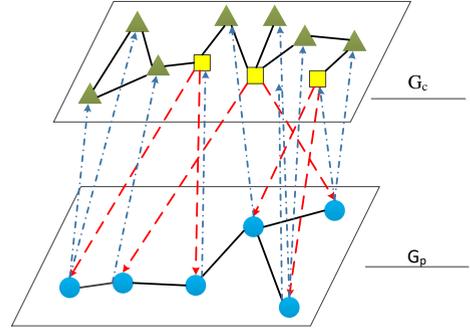

**FIG. 1** CPS coupling model for non-uniform power

If the node in the physical network fails, it cannot provide sufficient power to the cyber network. In the cyber network, important nodes and links will often have backup devices. If important node or link interruptions occur, their services can be transferred to the backup device. Therefore, important services can usually be recovered quickly, while other nodes that are not backed up will perform dynamic cyber distribution, which will be assigned to adjacent normal working nodes. This process may cause additional node failures as some nodes may undertake more cyber after redistribution than their assumed limits.

Studying the initial workload and capacity of the power network nodes shows that the number of cyber nodes that a power node can assume is limited. Therefore, the ball-slot model based on the infinite capacity of all power nodes in the existing literature [17] is not accurate and it is necessary to improve the ball-slot allocation method to construct a more reasonable coupling network.

Assuming that the maximum number of cyber nodes that a power node can support is related to its initial load, then:

$$N_z = \beta \cdot L_{nz}^{\mu} \quad (6)$$

The average maximum is:

$$<N_z> = \sum_{l_{nz}=0}^{\infty} P_d^z \cdot N_z = \beta \sum_{z=0}^{\infty} P_d^z \cdot L_{nz}^{\mu} \quad (7)$$

Assuming that the nodes in $G_p$ are marble bines, the nodes in $G_c$ are balls, and the sizes of $G_p$ and $G_c$ are $S_p$ and $S_c$, respectively, then the selectable position of the ball is $S_p<N_z>$, and $Pz_d$ represents the probability distribution of the degree of node. Thus, the probability of assigning the ball to the initial load $L_{nz}$ is:

$$P = \frac{P_d^z \cdot L_{nz}}{S_p \cdot <N_z>} \quad (8)$$

Defining the number of balls in the initial load $L_{nz}$ marble bin as the random variable $\omega$, then:

$$P(\omega = t) = \binom{l_{nz}}{t}\binom{S_c}{t} \cdot P^t \cdot (1-P)^{S_c}, t \leq l_{nz} \quad (9)$$

where $l_{nz}$ is the node degree of node $z$ in time step $n$. In addition, if the number of balls in a marble bin is defined as a random variable, then:

$$P(\omega' = t) = \sum_{z=0}^{\infty} P(\omega = t) \cdot P_d^z \qquad (10)$$

## III. RISK AREA PREDICTION MODEL FOR POWER CPS

The process of risk propagation can be considered as the development of a series of conditional probability events. The Markov process chain can describe the state change of these random events and the transition law between them. It can also predict the trend of events according to the transition probability between the initial state and the related state of the system. By combining the two, they can be used to describe and analyze the power CPS network risk propagation events with randomness and correlation.

### A. ANALYSIS OF RISK BROADCAST PROCESS CONSIDERING NETWORK INTERACTION

In a Markov process of a single network, the state of the system at each moment is only related to a state at a previous moment. By setting the state space of the system at time $i$ to $S_i$, the system needs $i$ time steps to transform from the initial state $S_0$ to $S_i$ through a series of state transitions, and the whole conversion process is combined by $\{C_1, C_2, ..., C_m\}$. Thus, the system comprehensive state transition probability is:

$$\begin{aligned} P_{C_1 C_2 \cdots C_m} &= P[C_1 = s_1] P[C_2 = s_2 | C_1 = s_1] \cdots \\ &\quad P[C_{m-1} = s_{m-1} | C_1 = s_1, C_2 = s_2, \cdots C_{m-1} = s_{m-1}] \\ &= P[C_1 = s_1] P[C_2 = s_2 | C_1 = s_1] \cdots \\ &\quad P[C_m = s_m | C_{m-1} = s_{m-1}] \end{aligned} \qquad (11)$$

However, for a power cyber-physical system composed of a cyber network and a physical network coupling, the total state space of the system is determined by the states of two networks, and one system node state change not only affects the neighboring nodes of the same network, but also the state of the nodes in the other network through the coupling component. The specific interaction process between system states is illustrated in Fig. 2.

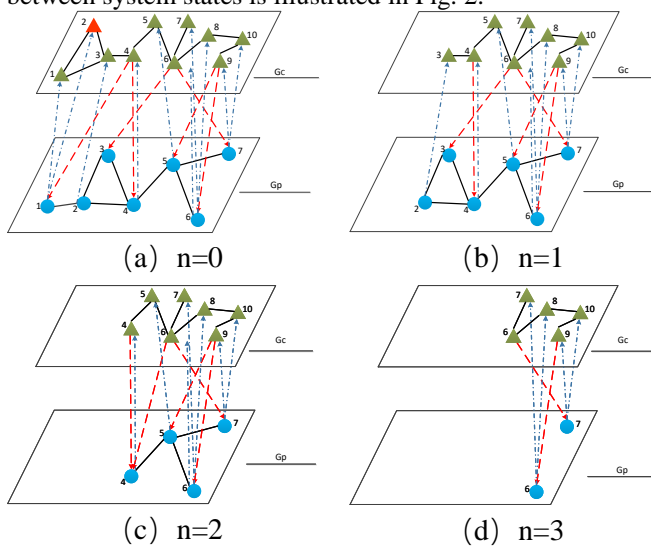

(a) n=0   (b) n=1
(c) n=2   (d) n=3

**FIG. 2** Status interaction process of cyber and physical network

The cyber system state space with time step n is represented by $S_{An}$ and the physical system state space of time step n is represented by $S_{Bn}$.

1) When n=0, there are ten cyber nodes in the cyber system state space $S_{a0}$, including one failed node 2, and six normal running physical nodes in the physical system state space $S_{b0}$. This system state is shown in Fig. 2(a).

2) When n=1, the state of the system node is judged. As node 2 in the cyber system is connected to the physical network node 1 through the coupling network, when cyber node 2 fails, the physical node 1 and its connected line are also affected and invalidated.

3) When n=2, the physical network sub-node fails and is removed, causing the failure of cyber node 3 through lost the power supply, and deleting the failed node and the corresponding line. This process is shown in Fig. 2(c).

4) When n=3 and above, the state of the dual-network system interacts and transforms until the network resumes its steady state.

It can therefore be observed that for the power cyber-physical coupling system, the system state at each moment may be related to multiple states at the previous moment and the single Markov chain probability framework is no longer applicable. In this work, two single heterogeneous Markov chains are combined to form a dependent Markov chain model, and the interactive transformation method is employed to capture the interdependence between the coupled networks. The cyber and physical network status and their corresponding transition probabilities are considered simultaneously when judging the system state space at a certain moment. The calculation formula of the system comprehensive state transition probability in the coupled network is then:

$$\begin{aligned} P_{C_1 C_2 \cdots C_m} &= P[C_1 = s_1] \{ P[C_2 = s_2 | C_{A1} = s_{A1}] \\ &\quad P[C_2 = s_2 | C_{B2} = s_{B2}] \} \cdots \\ &\quad \{ P[C_m = s_m | C_{Am-1} = s_{Am-1}] \\ &\quad P[C_m = s_m | C_{Bm} = s_{Bm}] \} \end{aligned} \qquad (12)$$

### B. RISK AREA PREDICTION MODEL BASED ON INTERDEPENDENT MARKOV CHAIN

#### 1) SYSTEM FULL STATE SPACE

The system full state space in the correlation Markov chain probability framework combines the cyber and physical network state space. The particularity of the interaction between the networks is taken into consideration and an auxiliary variable is also added to track the process of interaction of two network nodes between states. The resulting full state space $S_n$ of the power CPS system is characterized as follows:

$$S_n(X_n, I_n, Y_n, L_n, K_n)$$

in which $n$ is the total number of time steps for measuring the state change process during the risk propagation process.

For state variables of the power grid, $X_n$ is the physical network failure number, $I_n$ is the network stability index, 0 means it is unstable in the transition state, and 1 refers to stable operation in the absorption state. Additionally, $Y_n$ is the number of failures of the cyber, $L_n$ is the auxiliary variable, capturing the direction of transition in the cross-domain staggered framework, 0 means that the last faulty node is in the grid, and 1 means that it is in the cyber state. Finally, $K_n$ maps the history to the node of the last conversion through the quantization function and determines whether it is faulty.

### 2) STATE TRANSITION PROBABILITY OF DUAL NETWORK INTERACTION

The main reason for system state transition in the power cyber-physical coupling network is the load optimization reconfiguration caused by the node tolerance and the interaction of the dual network nodes through the coupling component. A state transition probability calculation method capable of capturing this relationship is proposed in this work. The recovery probability of the individual network is used to estimate the state transition probability of the whole system. First, the system state is divided into two categories: absorption state and transition state. Absorption state means that the load carried by all nodes in the network is below the constraint value. The system is in a stable running state at this time. Once this state is entered, the risk propagation is terminated and the system state is no longer changed. The transition state means that there is a node in the system that runs beyond the load constraint. Such a node may be cut off due to failure and the load it carries will be reassigned to its neighbors, triggering propagation of the node's risk state.

The probability that the physical network and the cyber state are restored to the absorption state from the transition state is represented by p(x) and q(y), respectively, where the variables $x$ and $y$ are the number of fault nodes exceeding the normal capacity in the physical network and the cyber, respectively. The degree of influence of the cyber state on the stable operation of the physical network is represented by the influence function $d$, and its expression is:

$$d: \{0,1,2,\ldots,m_C\} \to [0,1]$$

where $m_c$ represents the number of faulty component nodes in routers, switches, and cyber nodes in the cyber system. The range of d is between 0 and 1. The value from large to small corresponds to the impact of cyber system failure on the power grid from strong to weak.

The state transition process in the coupled network is described below based on the dependent Markov probability framework to construct the propagation dynamics equation. In this process, it is assumed that the initial failed node appears in the cyber, and the failure node transmission within the network during the load reconfiguration transfer process and the interaction between the networks in the coupling relationship are considered.

The system state transition probability at the next moment is obtained as follows:

1) When $K_n=0$, that is, the system as a whole is in the absorption state, the internal node will run stably without state transition, then the next time $K_{n+1}=0$, $S_{n+1}=S_n$, state transition probability $p(S_n \to S_{n+1})=1$;

2) When $K_n=1$, the system as a whole is in a transition state.

① If the new failed node appears in the cyber state at time n+1, the state transition probability expression is:

$$P(S_n \to S_{n+1}) = \begin{cases} q(y_n) & K_{n+1}=0 \\ 1-q(y_n) & K_{n+1}=1 \end{cases} \quad (13)$$

the risk continues to propagate and the system is still in the transition state;

② If the new failed node appears in the physical network along with the coupling network at time n+1, the state transition probability expression is:

$$P(S_n \to S_{n+1}) = \begin{cases} 1 - \dfrac{p(x_n)d(y_n)}{k_n + d(y_n)(1-k_n)} & K_{n+1}=0 \\ \dfrac{p(x_n)d(y_n)}{k_n + d(y_n)(1-k_n)} & K_{n+1}=1 \end{cases} \quad (14)$$

### 3) RISK AREA PREDICTION MODEL BASED ON INTERDEPENDENT MARKOV CHAIN

The system full state space obtained in the foregoing and the state transition probability under different conditions are combined in Eq. (11). At the same time, in order to simplify the system state, let $X(x, y)$ denote the asymptotic probability when the system state returns to the absorption state, $Y(x, y)$ represents the asymptotic probability when the system state is still in the transition state, and the following recursive equation is obtained by derivation:

$$X(x_i, y_i) = \alpha_1(x_i, y_i)X(x_{i-1}, y_i) + \alpha_2(x_i, y_i)X(x_{i-1}, y_{i-1}) + \alpha_3(x_i, y_i)X(x_{i-1}, y_{i-1}) \quad (15)$$

$$Y(x_i, y_i) = \alpha_4(x_i, y_i)X(x_{i-1}, y_i) + \alpha_5(x_i, y_i)Y(x_{i-1}, y_{i-1}) \quad (16)$$

where $x_i$ and $y_i$ represent the number of fault nodes of the physical network and the information network at time $i$, respectively, and other coefficients are:

$$\alpha_1(x_i, y_i) = \frac{p(x_i)(1-q(y_i))(1-p(x_{i-1}))}{p(x_{i-1})} \quad (17)$$

$$\alpha_2(x_i, y_i) = \frac{p(x_i)d(y_i)q(y_{i-1})(1-p(x_{i-1})d(y_{i-1}))}{p(x_{i-1})d(y_{i-1})} \quad (18)$$

$$\alpha_3(x_i, y_i) = (1-q(y_{i-1}))p(x_i)d(y_i)\left(p(x_{i-1}) - \frac{1-p(x_{i-1})}{d(y_{i-1})}\right) \\ + q(y_{i-1})p(x_i)(1-q(y_i))(1-d(y_i)) \quad (19)$$

$$\alpha_4(x_i, y_i) = \frac{(1-p(x_{i-1}))}{p(x_{i-1})} \quad (20)$$

$$\alpha_5(x_i, y_i) = q(y_{i-1})(1-p(x_{i-1})d(y_i)) - d(y_i)(1-p(x_{i-1})) \quad (21)$$

The power CPS risk area prediction model based on the dependent Markov chain is thus obtained. The model extracts and estimates the parameters by analyzing the historical data on the basis of capturing the interaction of the two networks, then calculates the probability of occurrence of the risk region containing different nodes using Eq. (16). The iteration number of the recursive equation is then taken as the order of failure of each node.

## IV. CROSS-ADAPTIVE GRAY WOLF OPTIMIZATION SOLUTION ALGORITHM

As is difficult to obtain effective prediction results by direct calculation using the recursive equation obtained above, a gray wolf optimization with fast convergence and parallelism is selected for the optimal solution of the problem. At the same time, because the solution domain of the risk region model has wide and multi-peak characteristics, the traditional gray wolf optimization method is prone to premature converge into local optimum in the process of solving this kind of model.

Due to the above-mentioned shortcomings, an adaptive grey wolf optimization based on crossover strategy is proposed in this study. The adaptive degree position adjustment strategy and cross optimal solution strategy are introduced to enhance the efficiency and accuracy of the grey wolf optimization result.

### A. DESCRIPTION OF ADAPTIVE GREY WOLF OPTIMIZATION FOR CROSSOVER STRATEGY

In the early stage of optimization of the grey wolf algorithm, it is difficult to control the search direction of the wolves. In addition, population diversity is prone to restriction, and the optimization results become limited to a single solution problem. Improvements to the strategy are proposed as follows:

1) Adaptive position adjustment strategy:

To determine the most suitable convergence speed, self-adaptive adjustment strategy is adopted to adjust the gray wolf position in the initial stage by comparing the current fitness value with the mean value of the wolf group fitness. The expression is as follows:

$$W(t+1) = \begin{cases} \dfrac{k_i W_1 + k_j W_2 + k_z W_3}{k_i + k_j + k_z} & k_n \geq k_{avg} \\ \dfrac{W_1 + W_2 + W_3}{3} & k_n < k_{avg} \end{cases} \quad (22)$$

where $W(t+1)$ represents the spatial azimuth coordinates of the gray wolf after the $t$-th iteration, t is the number of iterations, and $W_1$, $W_2$, and $W_3$ are the distances of the gray wolves whose current gray wolf positions are 1, 2, and 3 respectively. Additionally, $k_n$ and $k_{avg}$ respectively represent the current individual fitness and the reciprocal of the mean fitness, and $k_i$, $k_j$, and $k_z$ correspond to the reciprocal of the fitness of these grey wolves, respectively.

2) Cross-optimal solution strategy:

A cross-over strategy is proposed which uses the iteratively updated position-to-space coordinates to cross-correlate with the optimal population to screen out the location of the mutated population that leads to a single species of the population. The speed and position of each particle is determined using the following expression:

$$W_i'(t+1) = W_\mu(t) + |\gamma W_\mu(t) - W_i(t)| \cdot \beta \quad (23)$$

where $W_i(t+1)$ and $W_i(t)$ are the spatial coordinates of the individual wolves after resetting the local optimal solution, $W(t)$ is the current optimal solution, $\gamma$ is a random value in the range from the absolute value [1,2], and $\beta$ is a random value in the range [0,1], which is used to provide random fitness to define the attractiveness between the population and the prey.

### B. MODEL SOLUTION STEPS

The adaptive gray group optimization algorithm considering the crossover strategy works to add the fitness adjustment and the optimal solution set cross-correlation process to the approach of the original algorithm and to surround the prey. This process works to avoid the problem of random distribution of the population which can easily fall into the local optimum. The steps to solve the power CPS risk area prediction model using the adaptive grey wolf optimization algorithm of cross strategy are as follows:

1) Parameter initialization: The load and constraints are formally characterized according to the topology structure of the power CPS system and the tidal running parameters, and all nodes in the dual network operate normally.

2) Adaptive gray wolf initialization of the cross strategy: The individual size is set, the corresponding spatial position coordinates of the head wolf, the auxiliary wolf, and the subordinate wolf are selected, and the failure probability of each node is taken as the fitness value coefficient.

3) Simulate the system risk propagation process by using state transition probability: The initial information network fails node and the trigger system moves into a transition state. The initial system-wide state space is generated and the system's comprehensive state transition probability is calculated by considering the node load constraint, the degree of front-to-back state correlation, and the influencing factors of the coupled network.

4) Construct a risk-region prediction model for the dependent Markov chain: A risk area prediction model considering the specific attributes of the components and the operating state is established according to the constraints and the simulation process.

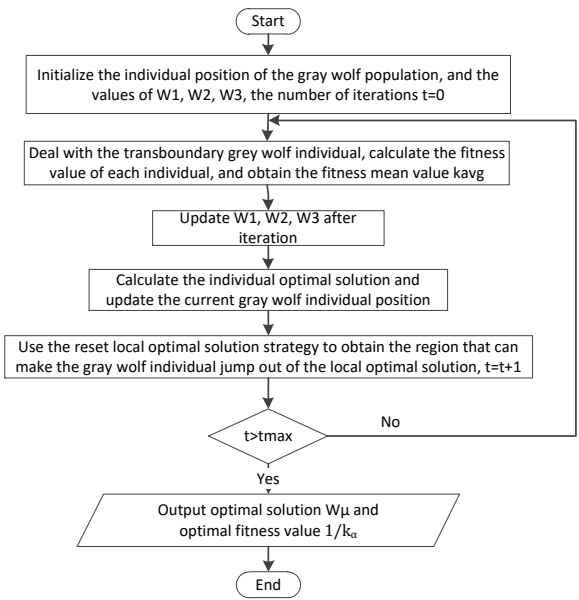

**FIG. 3** Flowchart of the cross-adaptive gray wolf optimization solution algorithm process

5) Describe the optimization process: The individual fitness value is calculated according to the constraint condition and the risk region transfer process. The fitness values of the head wolf, the auxiliary wolf, and the subordinate wolf are also determined. An adaptive degree position adjustment strategy and a cross optimal solution strategy are then introduced to update the current optimal individual and the global optimal individual.

6) Whether the end condition is satisfied is assessed, and if it is satisfied, the output of the global optimal solution is ended, otherwise jump to Step 3).

The cross-adaptive gray wolf optimization solution algorithm process is illustrated in Fig. 3

## V. SIMULATION EXPERIMENT AND DISCUSSION

### A. EXPERIMENTAL ENVIRONMENT DESIGN

The experimental environment was constructed by designing the physical layer, cyber layer, and the coupling relationship as the research object [6].

The physical layer builds an abstract network topology diagram based on the standard IEEE 39 bus system model and abstracts the generators, transformers, and other devices into physical nodes. The transmission lines between the devices are abstracted as edges, and the direction of the edges is not considered.

The cyber layer selection establishes a 110-node scale-free network based on the Barabasi-Albert model. The parameters are set as $m_0=3$, $m=2$, and the average node degree $<k>$ is 4.

The connection between the two layers employs the improved *bin and ball* allocation method in Section 1.3 to establish a non-uniform coupling connection. The line load factor is $\delta=\theta=\mu=2$ and the tolerance coefficient $\rho_c = \rho_p = 0.5$.

According to these parameters, a 149-node power cyber information physical system was built with IEEE 39-BA 110. The model was comprised of 39 physical nodes and 110 information nodes, 46 physical lines, 92 information lines, and 70 coupling branches. Some network structures are outlined as follows:

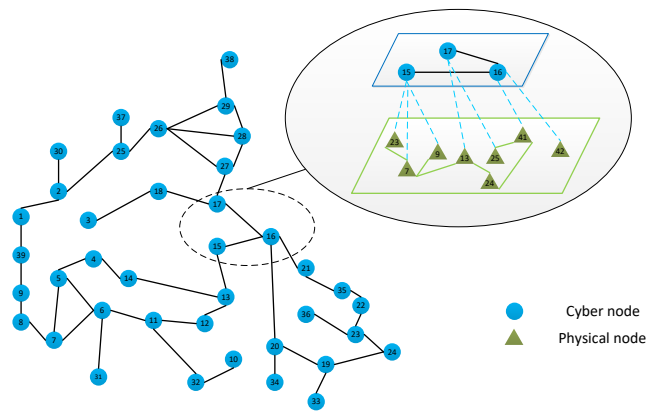

**FIG. 4** Node and line connection topology

### B. ANALYSIS OF EXAMPLES

The CPS coupling network IEEE 39-BA 110 was selected for analysis. To verify the effectiveness of the proposed prediction method, the following two steps were taken to analyze the example: 1) Power CPS risk predictive model validity analysis; 2) Effectiveness of the solution algorithm.

#### 1) POWER CPS RISK PREDICTIVE MODEL VALIDITY ANALYSIS

Based on the coupled system constructed in this paper, the fixed risk transfer probability model and the state transition probability model based on the dependent Markov chain were employed to predict the power CPS risk area. Table 1 lists some of the predictions obtained using the exhaustive method.

Table 1

PARTIAL PREDICTION PROBABILITY OF OCCURRENCE OF RISK AREA

| Risk area size (including the number of nodes) | | | Incidence probability | |
|---|---|---|---|---|
| Entire network | Information layer | Physical layer | Proposed model | Fixed risk transfer prediction model |
| 2 | 2 | 0 | 0.0289 | 0.0342 |
| 3 | 2 | 1 | $2.601\times10^{-3}$ | $6.332\times10^{-3}$ |
| 4 | 2 | 2 | $2.341\times10^{-3}$ | $4.171\times10^{-3}$ |
| | 3 | 1 | $3.422\times10^{-3}$ | |
| 5 | 3 | 2 | $3.98\times10^{-4}$ | $5.167\times10^{-4}$ |
| | 4 | 1 | $1.752\times10^{-4}$ | |
| 7 | 4 | 3 | $6.089\times10^{-6}$ | |
| | 5 | 2 | $4.172\times10^{-6}$ | $7.417\times10^{-6}$ |
| | 6 | 1 | $1.159\times10^{-6}$ | |
| 8 | 4 | 4 | $5.482\times10^{-6}$ | |
| | 5 | 3 | $4.035\times10^{-6}$ | $2.372\times10^{-6}$ |
| | 6 | 2 | $2.955\times10^{-6}$ | |
| | 7 | 1 | $0.097\times10^{-6}$ | |
| 9 | 5 | 4 | $6.437\times10^{-7}$ | |
| | 6 | 3 | $4.315\times10^{-7}$ | $5.538\times10^{-7}$ |
| | 7 | 2 | $1.731\times10^{-7}$ | |
| | 8 | 1 | $8.691\times10^{-8}$ | |

It can be seen from Table 1, for example, when the number of fault nodes is 7, the incidence probability obtained by fixed risk predication model is $7.417\times10^{-6}$, the results of proposed model can be divided into: 1) when the information layer has 4 failt nodes, the physical layer has 3 failt nodes, the incidence probability is $6.089\times10^{-6}$. 2) when the information layer has 5 failt nodes, the physical layer has 2 failt nodes, the incidence probability is $4.172\times10^{-6}$. 3) when the information layer has 6 failt nodes, the physical layer has 1 failt nodes, the incidence probability is $1.159\times10^{-6}$. According to the dependent Markov chain risk region prediction model, the probability of occurrence of risk regions of different scales can be calculated. By comparing the obtained results with two different transition probabilities, it can be determined that when the total number of nodes included in the risk region is the same, regardless of the distribution of the nodes in the information layer and the physical layer, the results obtained by using the fixed risk transition probability prediction are the same. However, in the process of actual system risk propagation, node failure caused by load constraint reconfiguration leads to changes in system network topology and dynamic power flow, which further affects the subsequent risk propagation trend. The fixed risk transfer probability ignores the correlation between the state of the component nodes, so a deviation of the prediction results will occur. This effect is not obvious when the risk area is small, but with increasing scale, this effect becomes more apparent. Therefore, the prediction model in this paper can distinguish between the different network node regions, which is beneficial to obtaining more accurate prediction results.

Analysis of the results of state transition probability calculation also shows that the probability of occurrence of the 8-node risk region with four physical nodes and four information nodes is greater than that of some 7-node risk regions. The probability of occurrence of the scale risk area is mainly due to the strong correlation between some nodes in the coupled network. Once one of the nodes fails, the other will inevitably fail.

2) SOLUTION ALGORITHM VALIDITY AND EFFICIENCY ANALYSIS

The five model solution algorithms of ant colony optimization (ACO), artificial bee colony algorithm (ABC), gravitational search algorithm (GSA), grey wolf optimization (GWO), and cross-adaptive grey wolf optimization (CAGWO) were selected for comparative experiments. The algorithm parameters were set as shown in Table 2.

Table 2
PARAMETER SETTINGS OF FIVE ALGORITHMS

| Solution algorithm | $N$ | $K_{max}$ | $G_0$ | $L_{max}$ | $\beta$ | $\omega$ | $\eta$ |
|---|---|---|---|---|---|---|---|
| ACO | 120 | 1000 | —— | 100 | —— | —— | —— |
| ABC | 120 | 1000 | —— | 100 | —— | —— | —— |
| GSA | 120 | 1000 | 100 | —— | 20 | —— | —— |
| GWO | 40 | 1000 | —— | —— | —— | 8 | 2 |
| SAGWO | 40 | 1000 | —— | —— | —— | 8 | 2 |

The network functional evaluation of the prediction results obtained by the five solution algorithms was carried out by using the remaining network functional evaluation indicators. The evaluation indicators are the maximum connectivity rate and load loss of the network.

● Maximum network connectivity

The maximum connectivity of the network is an important indicator for measuring the connectivity of the coupled network. It refers to the probability that one edge is randomly selected in the network after the predicted risk region is removed. This edge belongs to the largest connected domain. In a coupled network with interdependence, each node is connected to each other. After the risk area is removed, the load reconfiguration is performed after the m time step, and when the network resumes the absorption state, the maximum functional area connected to the network is interdependent.

The number of edges is denoted by R, then the maximum network connectivity $R_{max}$ is:

$$R_{\max}=\frac{\sum_{i\in(V_{c-normal}\cup V_{p-normal})}R}{V(V-1)} \quad (24)$$

● Load loss indicator

The load loss degree index refers to the degree of load loss of the network after the risk area is removed, and is used to evaluate the importance of the risk area. The expression is:

$$\eta = \frac{\sum_{i \in V_{p\text{-}normal}} L^p_{(n+1)i} - \sum_{i \in V'_{p\text{-}normal}} L^p_{ni}}{\sum_{i \in V'_{p\text{-}normal}} L^p_{ni}} \quad (25)$$

where $LP_{(n+1)i}$ and $LP_{ni}$ are the physical node loads during the time step n+1 risk area excision and time step n risk area excision, respectively.

Prediction results of the five algorithms are provided for comparison, as shown in Fig. 5(a) and (b).

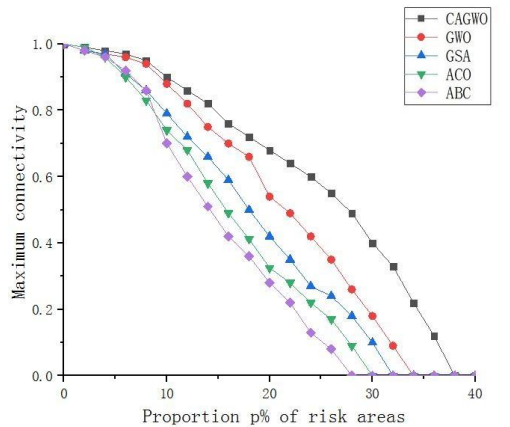

(a) Maximum connectivity of the network

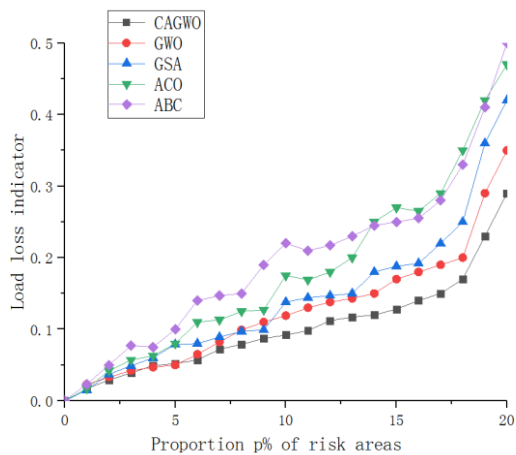

(b) Load loss indicator

**FIG. 5** Functional evaluation of residual network of power CPS

Figure 5(a) and (b) can be used to conduct a functional evaluation of the remaining CPS network after the risk area is removed. When the risk area is small, the maximum connectivity of the remaining networks obtained by the five algorithms is similar. Because the risk spread range is small at this time, there are many remaining network work nodes, so structural connectivity is not significantly diminished. As the proportion of risk areas increases, the results of the five algorithms are more obvious. The remaining network functionality obtained by the algorithm used in this paper is significantly superior to other algorithms. The prediction results provide the least harm to the remaining network and can retain the maximum connectivity and minimize load loss of the remaining network.

The optimization time of different algorithms for different CPS scale is compared in Fig. 6.

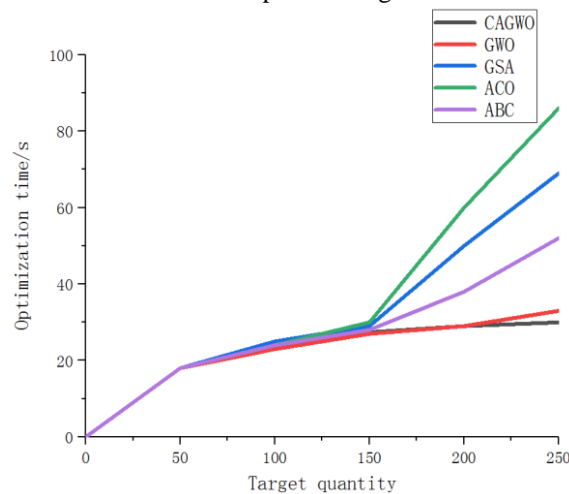

**FIG. 6** Comparison of solving time of different scale tasks

When the power CPS is small, the calculation time of the five algorithms is similar, indicating that the small scale increases the CPS scale and the number of nodes is large. In summary, example 1) verifies the superiority of the model used in this paper, which can distinguish between the different network node regions, getting more accurate predictions. It can be seen from example 2), the solution algorithm proposed in this paper can adaptively adjust gray wolf fitness. The local precocity convergence characteristics are changed, the optimization results are quickly converged, and the time-consuming aspect presents a significant advantage. This means that when the algorithm is applied to a large number of nodes of power CPS, the prediction model can be solved more efficiently.

## VI. CONCLUSION

A power CPS risk region prediction model based on dependent Markov chain was proposed in this paper which can distinguish node heterogeneity and capture the interdependencies between the physical network and cyber network. An improved cross-adaptive grey wolf optimization was then used to solve this model, with the ultimate goal of predicting the probability of the occurrence of a risk area. This paper provides a theoretical reference for preventing the spread of risks in real world applications.

Future work will consider different information network topologies and types of transmission services to further improve the risk area prediction model. Another interesting topic for future study would be to extend this work to potential applications in risk region prediction of an integrated energy system [23].

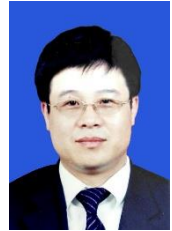

**FFZHAOYANG QU** received a PhD degree in electrical engineering from China Northeast Electric Power University in 2010, and his M.S. degree in Dalian University of Technology in 1988. He is currently a professor in the School of Information Engineering of Northeast Electric Power University. His interests include smart grid and power information, virtual reality, network technology.

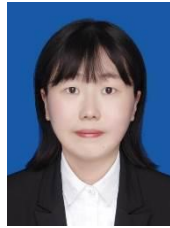

**QIANHUI XIE** received a B.S degree in electrical engineering from the University of South China in 2013 and 2017, and an M.S degree at Northeast Electric Power University. Her interests include power cyber-physical systems, smart grid and power information, and smart electricity.

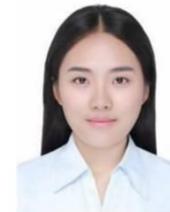

**YUQING LIU** is currently pursuing a PhD degree in electrical engineering at the University of Bath. Her research interest is ionospheric changes caused by natural aurora and lightning.

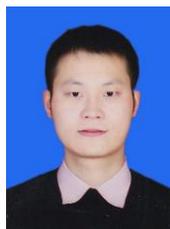

**YANG LI** was born in Nanyang, China. He received a PhD degree in electrical engineering from North China Electric Power University, Beijing, China in 2014. He is currently an Associate Professor with the School of Electrical Engineering, Northeast Electric Power University, Jilin, China. He is also a China Scholarship Council funded Post-Doctoral Researcher at the Argonne National Laboratory, Lemont, IL, USA. His research interests include power system stability and control, integrated energy system, renewable energy integration, and smart grids. He is an Associate Editor of IEEE Access.

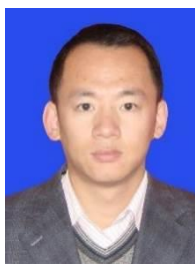

**LEI WANG** was born in Jilin, China. He is pursuing a PhD degree in electrical engineering at Northeast Electrical Power University. He is currently an Associate Professor with the School of Information Engineering and his research interest is information processing in smart grid.

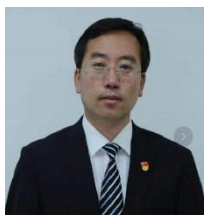

**PENGCHENG XU** received a B.S. degree from Changchun University of Technology. He is currently a Senior Engineer at State Grid Jilin Power Co., Ltd. Siping Power Supply Company. His research interest is power system automation.

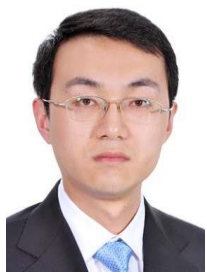

**YUGUANG ZHOU** was born in May 1980. He graduated from Shanghai Jiao Tong University with a master's degree in power systems and automation, and is now working in the Power Dispatching Control Center of the State Grid Jilin Electric Power Co., Ltd. His research direction is power system dispatching automation.

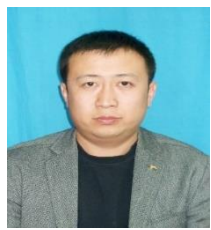

**JIAN SUN** received a B.S. degree from Northeast Electric Power University in 2001. He is currently a Senior Engineer at State Grid Jilin Power Co., Ltd. Baishan Power Supply Company. His research interest is safety monitoring in the smart grid.

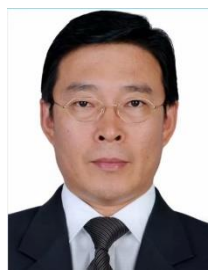

**KAI XUE** received an M.S. degree from Northeast Normal University. He is currently a Senior Engineer at State Grid Jilin Power Co., Ltd. Training Center. His research interest is power system automation.

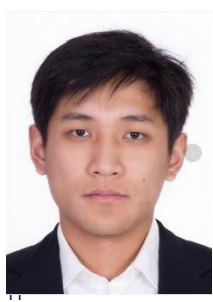

**MINGSHI CUI** received a B.S. degree from North China Electric Power University in 2010. He is currently an Engineer at State Grid Inner Mongolia Eastern Electric Power Company. His research interest is information processing in the smart grid.